\documentclass{elsart}

\usepackage{amssymb,amsmath,mathrsfs,eufrak,mathabx}

\newtheorem{exm}{Example}[section]

\newcommand{\ar}{\rightarrow}
\newcommand{\la}[1]{\it {#1}\rm}

\newcommand{\mA}{\mathcal{A}}

\newcommand{\mZ}{\mathbb{Z}}
\newcommand{\mN}{\mathbb{N}}
\newcommand{\mI}{\mathcal{I}}

\newcommand{\mO}{\mathcal{O}}

\newcommand{\mC}{\mathfrak{C}}
\newcommand{\mS}{\mathcal{S}}

\newcommand{\df}[1]{\begin{defn} #1 \end{defn}}

\newcommand{\te}[1]{\begin{thm} #1 \end{thm}}
\newcommand{\ex}[1]{\begin{exm} #1 \end{exm}}
\newcommand{\co}[1]{\begin{cor} #1 \end{cor}}
\newcommand{\rk}[1]{\begin{rem} #1 \end{rem}}

\newcommand{\bb}[1]{\mathbb{#1}}

\newcommand{\sq}{\sqbullet}

\begin{document}

\begin{frontmatter}



\title{A New Approach to Abstract Machines \\ - Introduction to the Theory of \\  Configuration Machines \\ }
\author{Zhaohua Luo}

 \ead{zluo@azd.com}
\ead[url]{http://www.algebraic.net/cag}

\begin{abstract}

An abstract machine is a theoretical  model designed to perform a
rigorous study of computation. Such a model usually consists of
configurations, instructions, programs, inputs and outputs for the
machine. In this paper we formalize these notions as a very simple
algebraic system, called a configuration machine. If an abstract
machine is defined as a configuration machine consisting of
primitive recursive functions then the functions computed by the
machine are always recursive. The theory of configuration machines
provides a useful tool to study universal machines.

\end{abstract}

\end{frontmatter}


\section{Introduction}
An abstract machine is a theoretical  model designed to perform a
rigorous study of computation. Such a model usually consists of
detailed descriptions of how to define configurations, instructions,
programs, inputs and outputs for the machine, and how to perform
computations step by step. In this paper we formalize these notions
as a very simple algebraic system, called a \la{configuration
machine}. If an abstract machine is defined as a configuration
machine which consists of only primitive recursive functions then
the functions computed by the machine are always recursive. This
fundamental fact is usually justified by Church thesis but is rather
tedious to prove in the traditional approach. The theory of
configuration machines provides a useful tool to study universal
machines.

\section{Configuration Machines}

Let $\mN$ be the set of natural numbers. A function from $\mN^k$ to
$\mN$ for any $k \ge 0$ is simply called a \la{function}. The
\la{class of primitive recursive functions} is the smallest class of
total functions including the initial functions and closed under
composition and primitive recursion. The smallest class of partial
functions that includes the initial functions and is closed under
composition, primitive recursion, and the $\mu$ operator is called
the \la{class of recursive functions}. The reader is refer to
\cite{odi}~\cite{rog}~\cite{shoe1} for the general theory of
recursive functions.

We say a function $f: \mN^k \ar \mN^n$ is \la{primitive recursive}
if each component $f_i = p_i \cdot f: \mN^k \ar \mN$ $(1 \le i \le
n)$ is so, where $p_i: \mN^n \ar \mN$ is the $i$-th projection.

Let $\mN^{\infty} = \mN^1 \cup \mN^2 \cup \mN^3 \cup ..\cup \mN^k
\cup ....$. A function $\mI: \mN^{\infty} \ar \mN^n$ is called
\la{primitive recursive} if the restriction $\mI^k$ of $\mI$ on each
$\mN^k \subset \mN^{\infty}$ is primitive recursive.

A \la{configuration computer} (or \la{$C$-computer})  $\mC =
(\bb{C}, \bb{S}, \bb{P}, \bb{M}, 0, \tau, \circ, \bullet)$ is an
algebraic system consisting of four nonempty sets $\bb{C}, \bb{S},
\bb{P}, \bb{M}$, an element $0 \in \bb{P}$, and three functions
$\tau: \bb{C} \ar \bb{S}$, $\circ: \bb{P} \times \bb{C} \ar \bb{C}$,
$\bullet: \bb{M} \times \bb{S} \ar \bb{P}$ such that

C1. $0 \circ c = c$ for any $c \in \bb{C}$.

C2. For any $e \in \bb{M}$ there is only finitely many $s \in
\bb{S}$ with $e \bullet s \ne 0$.

A \la{$C$-machine} is a $C$-computer $\mC$ together with two
functions $\mI: \mN^{\infty} \ar \bb{C}$ and $\mO: \bb{C} \ar \mN$,
called the \la{input and output} for $\mC$.

A \la{program} for a $C$-computer $\mC$ is a function $m: \bb{S} \ar
\bb{P}$ such that there is only finitely many $s \in \bb{S}$ with
$m(s) \ne 0$. Each $e \in \bb{M}$ determines a program $\nu(e)$ such
that $\nu(e)(s) = e \bullet s$. Denote by $Prog(\mC)$ the set of
programs for $\mC$. We have a function $\nu: \bb{M} \ar Prog(\mC)$.
We say $\mC$ is \la{complete} if $\nu$ is bijective. We say $\mC$ is
\la{canonical} if $\bb{M} = Prog(\mC)$ and $m \bullet s = m(s)$ for
any $m \in Prog(\mC)$ and $s \in \bb{S}$.

Elements in $\bb{C}, \bb{S}, \bb{P}$ are called \la{configurations,
situations and instructions} respectively.

A configuration $c \in \bb{C}$ is called \la{terminal for} an
element $e \in \bb{M}$ (resp. a program $m$) if $e \bullet \tau(c) =
0$ (resp. $m(\tau(c)) = 0$).

Suppose $\mC$ is a $C$-computer. We define a sequence of functions:

total function $\rho: \bb{M} \times \bb{C} \ar \bb{C}$,

total function $\delta: \bb{M} \times \bb{C} \times \mN \ar \bb{C}$,

partial function $h: \bb{M} \times \bb{C} \ar \bb{C}$,

partial function $\varphi: \bb{M} \times \bb{C} \ar \bb{C}$.

Suppose $e \in \bb{M}, c \in \bb{C}$. Define:

$\rho(e, c) = (e \bullet \tau(c)) \circ c$,

$\delta: \bb{M} \times \bb{C} \times \mN \ar \bb{C}$ inductively as
$\delta(e, c, 0) = c$ and $\delta(e, c, n+1) = \rho(e, \delta(e, c,
n))$ for any $n \ge 0$.

Let $h(e, c) = \mu i(e \bullet \tau(\delta(e, c, i)) = 0)$, i.e.
$h(e, c)$ is the least number $i$ such that $\delta(e, c, i)$ is
terminal for $e$ if such an $i$ exists, otherwise $h(e, c)$ is
undefined.

Let $\varphi(e, c) = \delta(e, c, h(e, c))$.

Suppose $\mC$ is a $C$-machine.

For any $k > 0$ let $\psi^k: \bb{M} \times \mN^k \ar \mN$ be the
partial function defined by $\psi^k(e, a_1, ..., a_k) =
\mO(\varphi(e, \mI(a_1, ..., a_k)))$ for any $(a_1, ..., a_k) \in
\mN^k$.

For any $e \in \bb{M}$ and $k > 0$ let $\phi_e^k: \mN^k \ar \mN$ be
the partial function such that $\phi_e^k(a_1, ..., a_k) = \psi^k(e,
a_1, ..., a_k)$.

We point out that the definition of $\phi_e^k$ is completely
determined by the program $\nu(e)$, because $\rho(e, c) =
\nu(e)(\tau(c)) \circ c$, and $c$ is terminal for $e$ iff
$\nu(e)(\tau(c)) = 0$. Thus if $e$ and $e'$ determine the same
program, then $\phi_e^k = \phi_{e'}^k$.

For any program $m$ for $\mC$ we define a partial function
$\phi_m^k$ as above, using $\rho(e, c) = m(\tau(c)) \circ c$  and
$m(\tau(c)) = 0$ as the criterion for a terminal configuration for
$m$.

Let $\mathscr{C}_k = \{\phi_e^k \ | \ e \in \bb{M}\}$.

Let $\mathscr{C} = \mathscr{C}_1 \cup \mathscr{C}_2 \cup
\mathscr{C}_3 \cup ...$

A partial function $f: \mN^k \ar \mN$ is called
\la{$\mC$-computable} if $f \in \mathscr{C}$.

An \la{arithmetic $C$-machine} is a $C$-machine $\mC$ such that the
following conditions are satisfied:

(1) $\bb{C} = \mN^r$ for some $r > 0$.

(2) $\mN = \bb{S} = \bb{P} = \bb{M}$.

(3) All the functions $\tau, \circ, \bullet, \mI, \mO$ are primitive
recursive.

Since all the functions involved in the definition of $\psi$ and
$\psi^k$ are recursive for an arithmetic machine, we have the
following fundamental theorem:

\te{\label{main-theorem} Suppose $\mC$ is an arithmetic $C$-machine.
Then

1. For any $k > 0$ the partial function $\psi^k$ is recursive.

2. For any number $e$ and $k > 0$ the partial function $\phi_e^k$ is
recursive.

3. For any program $m$ and $k
> 0$ the partial function $\phi_m^k$ is recursive (since $m$ is primitive recursive).

4. Any $\mC$-computable function is recursive.}

A partial $k+1$-ary recursive function $F(y, x_1, ..., x_k)$ from
$\mN^{k+1}$ to $\mN$ is called a \la{universal function} for $k$-ary
recursive functions if for any partial $k$-ary function $f(x_1, ...,
x_k)$ from $\mN^k$ to $\mN$ there exists a number $y \in \mN$ such
that $f(x_1, ..., x_k) = F(y, x_1, ..., x_k)$ for any $(x_1, ...,
x_k) \in \mN^k$.

A \la{universal machine} is a $C$-machine $\mC$ such that $\bb{M} =
\mN$ and for any $k
> 0$ the partial function $\psi^k: \mN^{k+1} \ar \mN$ is universal for $k$-ary
recursive functions

If $\mC$ is arithmetic then $\psi^k$ is recursive by
Theorem~\ref{main-theorem}. Thus $\mC$ is universal if any recursive
function is $\mC$-computable. Hence we have the following

\te{\label{package} An arithmetic $C$-machine $\mC$ is universal if
any recursive function is $\mC$-computable. }

We shall see in Section~\ref{counter} and \ref{turing} that the
$C$-machines determined by Turing machine and unlimited register
machine are universal.

In practice a concrete $C$-machine is often given as a
\la{configuration algebra} (plus input and out functions) which is
defined as follows:

\df{A configuration algebra $\mA = (\bb{C}, \bb{S}, \bb{P}, \tau,
\circ)$ is an algebraic system consisting of three nonempty sets
$\bb{C}, \bb{S}, \bb{P}$ and two functions $\tau: \bb{C} \ar \bb{S}$
and $\circ: \bb{P} \times \bb{C} \ar \bb{C}$. A configuration
algebra is arithmetic if $\bb{C} = \mN^r$ for some $r > 0$,  $\bb{S}
= \mN$ and $\bb{P} = \mN^+$, where $\mN^+$ is the set of positive
integers.}

By a \la{program} for a configuration algebra $\mC$ we mean a
function from a finite subset of $\bb{S}$ to $\bb{P}$. Let $\bb{M}$
be the set of all programs for $\mC$. Assume $0 \notin \bb{P}$ and
let $P^* = P \cup \{0\}$.  For any program $m \in \bb{M}$ we define
$m \bullet s = m(s)$ if $m(s)$ is defined and $m \bullet s  = 0$
otherwise. Then $\mC^* = (\bb{C}, \bb{S}, \bb{P}^*, \bb{M}, 0, \tau,
\circ, \bullet)$ is a canonical $C$-machine, and any complete
$C$-machine arises in this way, up to an isomorphism of $\bb{M}$.
Conversely, any $C$-machine determines a configuration algebra by
forgetting $\bb{M}$, $\bullet$ and changing $\bb{P}$ to $\bb{P} -
0$. Thus the notions of configuration algebra and complete or
canonical  $C$-computer are equivalent.

\ex{The Turing algebra is a configuration algebra defined as
follows. Let $S = \{s_0, s_1, ...\}$ be an infinite set of symbols.
A two-way $S$-tape is a function from the set $\mZ$ of integers to
$S$ sending all but finitely many integers to $s_0$. Suppose $u$ is
such a tape. If we change $u(0)$ to $s \in S$ we obtain a new tape
denoted by $S_su$. Let $Ru$ and $Lu$ be the tapes such that $(Ru)(i)
= u(i+1)$ and $(Lu)(i) = u(i-1)$ respectively. Let $Tape_S$ be the
set of all $S$-tapes. Let $N = \{q_0, q_1, ...\}$ be a set of
states. Let $\bb{C} = N \times Tape_S$, $\bb{S} = N \times S$,
$\bb{P} = (N \times (S \cup \{R, L\})$. If $u$ is a tape and $q_i$
is a state let $\tau(q_i, u) = (q_i, u(0))$. Define $(q_i, s) \circ
(s_j, v) = (q_i, S_sv)$, $(q_i, R) \circ (q_k, v) = (q_i, Rv)$ and
$(q_i, L) \circ (q_k, v) = (q_i, Lv)$ (cf.
~\cite{dav1}~\cite{dav2}~\cite{turing}). }

\ex{The unlimited register algebra is a configuration algebra
defined as follows. A one-way tape is an infinite sequence $u =
(u_0, u_1, ...)$ of numbers such all but a finite number of
components are not $0$. Denote by $Tape$ the set of one-way tapes.
Let $\bb{C} = \mN \times Tape$, $\bb{S} = \mN$ and $\tau(a, u) = a$.
Let
\[\bb{P} = \{Z(r), S(r), T(r, s), J(r,s, t)\}_{r, s, t \in \mN}.\]
Define

$Z(r) \circ (a, u) = (a+1, (u_0, ..., u_{r-1}, 0, u_{r+1}, ...))$.

$S(r) \circ (a, u) = (a+1, (u_0, ..., u_{r-1}, u_r + 1, u_{r+1},
...))$.

$T(r, s) \circ (a, u) = (a + 1, (u_0, ..., u_{s-1}, u_r, u_{s+1},
...))$.

$J(r, s, t) \circ (a, u) = (t, u)$ if $u_r = u_s$ and $J(r, s, t)
\circ (a, u) = (a+1, u)$ otherwise (cf.~\cite{cut}). }

\section{\label{finite-function} Finite Functions}

A \la{$0$-finite function} is a function $u: \mN \ar \mN$ sending
all but a finitely many numbers to $0$ (or equivalently, a function
from a finite subset of $\mN$ to $\mN^+$); a $0$-finite function can
also be viewed as an infinite sequence $(a_0, a_1, ...)$, with
almost all components are zero, called a \la{$0$-finite sequence}.
Denote by $\Omega_{\mN}$ the set of all $0$-finite functions. Let
$\bullet: \Omega_{\mN} \times \mN \ar \mN$ be the canonical function
such that $u \bullet x = u(x)$ for any finite $0$-function $u$ and
number $x$.

The set $\Omega_{\mN}$ of all $0$-finite functions plays an
important role in our algebraic approach to the theory of abstract
machines. A crucial fact about $\Omega_{\mN}$ is that it is
effectively denumerable, i.e. there is an effectively computable
bijection from $\Omega_{\mN}$ to $\mN$. In this paper we shall use
the bijection determined by a \la{primitive pairing}.

By a \la{primitive pairing} we mean an algebra $(\mN, \alpha, \beta,
[ \ ])$ where $\alpha, \beta: \mN \ar \mN$ and $[ \ ]: \mN \times
\mN \ar \mN$ are primitive recursive functions such that the
following conditions are satisfied for any integer $a, b \ge 0$:

1. $\alpha[a, b] = a$.

2. $\beta[a, b] = b$.

3. $[\alpha a, \beta a] = a$.

4. $[0, 0] = 0$.

5. $\alpha z \le z$ and $\beta z < z$ for any $z > 0$.

We shall write $[a_1, a_2, ..., a_n]$ for $[a_1, [a_2, [a_3, ...,
[a_{n-1}, a_n]...]]]$.

 \rk{A primitive pairing is a genoid in the category of
sets in the sense of~\cite{luo}. }

\ex{The Cantor pairing function given by \[[a, b] = 1 + 2 + ... + (a
+ b) + b = 1/2(a+b)(a + b + 1) + b\] defines a primitive pairing:
\[
\underbrace{[0, 0]}_0, \underbrace{[1, 0]}_1, \underbrace{[0, 1]}_2,
\underbrace{[2, 0]}_3, \underbrace{[1, 1]}_4, \underbrace{[0, 2]}_5,
\underbrace{[3, 0]}_6, \underbrace{[2, 1]}_7, \underbrace{[1, 2]}_8,
\underbrace{[0, 3]}_{9}, ...\]}

\rk{Note that the condition 5 is not symmetric for $\alpha$ and
$\beta$. For Cantor pairing we have $\alpha 1 = 1 \le 1$ and $\beta
1 = 0 < 1$. Thus the pairing
\[[a, b]^* = 1 + 2 + ... + (a + b) + a = 1/2(a+b)(a + b + 1) + a\] is
not a primitive pairing: instead of $1 = [1, 0, ..., 0]$ the pairing
$[ \ ]^*$ gives us $1 = [0, 0, ..., 1]^*$, which means there is no
$0$-finite sequence determined by $1$.}

In the following we shall fix a primitive pairing.

Suppose $a$ is any number. For any number $n$ define $\alpha_n(a) =
\alpha\beta^na$, which is called the \la{$n$-th component} of $a$.
Then
\[\alpha_0(a) = \alpha(a),\]
\[\alpha_1(a) = \alpha\beta (a),\]
\[a = [\alpha_0(a), \alpha_1(a), ..., \alpha_n(a),
\beta^{n+1}a].\]

\bf{Notation:} \rm Sometimes it is convenient to write $(a)_n$ for
$\alpha_n(a)$, and write $(a)_{n, m}$ in place of
$\alpha_m(\alpha_n(a))$, etc. Then we have
\[a = [(a)_0, (a)_1, ..., (a)_n, \beta^{n+1}(a)].\]

Note that \[[(a)_0, (a)_1, ..., (a)_n, 0] = [(a)_0, (a)_1, ...,
(a)_n, 0, 0] = [(a)_0, (a)_1, ..., (a)_n, 0, 0, 0, ..., 0].\] If
$(a_0, a_1, ...) \in \Omega_{\mN}$ then we write $[a_0, a_1, ...]$
for $[a_0, a_1, ..., a_k, 0]$, where $k $ is any number such that
$a_i = 0$ for $i > k$.

By induction one can show that $\alpha_aa = (a)_a = \beta^aa = 0$.
Thus $\alpha_n(a) = (a)_n = 0$ for any $n \ge a$. Hence $((a)_0,
(a)_1, ...)$ has only finitely many nonzero components for any
number $a$, i.e. $((a)_0, (a)_1, ...) \in \Omega_{\mN}$. We have
\[a = [(a)_0, (a)_1, ...., (a)_a, 0].\]
Thus it is  legitimate
 to write $a = [(a)_0, (a)_1, ...]$ for any number
$a$.

We shall write $a \sq b$ for $(a)_b$. The function $\sq: \mN \times
\mN \ar \mN$ is a primitive recursive function.

If $f$ is a $0$-finite function let $\mS(f) = [f(0), f(1), ...]$,
Then we have $f(x) = \mS(f) \sq x$. Conversely any number $a$
determines a $0$-finite function (or sequence) $((a)_0, (a)_1, ...)$
and $\mS((a)_0, (a)_1, ...) = a$. Thus we obtain an effectively
computable bijection $\mS: \Omega_{\mN} \ar \mN$.

The binary operation $\sq$ is called a \la{universal operation (or
function)} for $0$-finite functions.

We summarize the main properties concerning $\sq$ in the following
theorem:

\te{There is a pair of functions $\sq: \mN \times \mN \ar \mN$ and
$[x_0, x_1, ...]: \Omega_{\mN} \ar \mN$ such that the following
conditions are satisfied:

1. For any $a \in \mN$ the sequence $(a \sq 0, a \sq 1, ...)$ is
$0$-finite.

2. $[a_0, a_1, ...] \sq i = a_i$ for any $(a_0, a_1, ...) \in
\Omega_{\mN}$.

3. $[a \sq 0, a \sq 1, ...] = a$.

4. $a \sq b = 0$ if $b \ge a$.

5. $\sq$ is primitive recursive.

6. Let $\beta: \mN \ar \mN$ be the function such that $\beta(a) = [a
\sq 1, a \sq 2, ...]$. Then $\beta$ is primitive recursive.

7. For any $0$-finite function $f$ there is a unique number $e$ such
that $e \sq x = f(x)$ for any number $x$.}

\section{\label{counter}Counter Machines}

A \la{counter machine} is an arithmetic $C$-machine satisfying the
following conditions:

1. $\bb{C} = \mN \times \mN$, $\bb{S} = \bb{P} = \bb{M} = \mN$; a
congruence $(a, u) \in \bb{C}$ consists a number in the counter and
a code $u$ for a one-way tape.

2. $\tau: \bb{C} \ar \mN$ is the projection sending $(a, u)$ to $a$.

3. $\bullet = \sq$.

4. $\mI^k(a_1, a_1, ..., a_k) = (0, [a_1, ..., a_k, 0, ...])$ and
$\mI = \mI^1 \cup \mI^2 \cup \mI^3 \cup ...$.

5. $\mO(a, u) = (u)_0$.

Thus a counter machine reduces to a single primitive recursive
function $\circ: \mN \times \bb{C} \ar \bb{C}$ such that $0 \circ
(a, u) = (a, u)$ for any configuration $(a, u)$.

Applying Theorem \ref{main-theorem} to a counter machine we have the
following

\te{If $\mC$ is a counter machine then any $\mC$-computable function
is recursive. }

We define two types of counter machines (see
~\cite{boolos}~\cite{cut}~\cite{shoe2}):

Suppose $r, s, t \ge 0$.

Unlimited Register Configuration Machine (URCM):

If $p = 4r + 1$ let $p \circ (a, u) = (a+1, [(u)_0, ..., (u)_{r-1},
0, (u)_{r+1}, ...])$.

If $p = 4r + 2$ let $p \circ (a, u) = (a+1, [(u)_0, ..., (u)_{r-1},
(u)_r+1, (u)_{r+1}, ...])$.

If $p = 4[r, s] + 3$ let $p \circ (a, u) = (a+1, [(u)_0, ...,
(u)_{s-1}, (u)_r, (u)_{s+1}, ...])$.

If $p = 4[r, s, t] + 4$ let $p \circ (a, u)  = (t, u)$ if $(u)_r =
(u)_s$, and $p \circ (a, u) = (a +1, u)$ otherwise.

Abacus Configuration Machine (ACM):

If $p = 2[r, s] + 1$ let $p \circ (a, u) = (s, [(u)_0, ...,
(u)_{r-1}, (u)_r +1, (u)_{r+1}, ...])$.

If $p = 2[r, s, t] + 2$ let $p \circ (a, u)  = (s, [(u)_0, ...,
(u)_{r-1}, (u)_r -1, (u)_{r+1}, ...])$ if $(u)_r \ne 0$, and $p
\circ (a, u) = (t, u)$ otherwise.

The operations $\circ$ defined above are primitive recursive
(Definition by Cases, cf~\cite{cut}, p.37). Thus URCM and ACM are
counter machines.

For the unlimited register machine (URM) defined in \cite{cut} there
are four classes of instructions: $Z(n), S(n), T(m, n)$ and $J(m, n,
q)$ with $m, n, q \ge 1$. A finite list $P = I_1, ..., I_d$ of these
instructions constitutes a program for URM. Define $code(Z(n)) =
4(n-1) + 1$, $coed(S(n)) = 4(n-1) + 2$, $code(T(m, n,)) = 4[m-1,
n-1] + 3$ and $code(J(m, n, q)) = 4[m-1, n-1, q-1] + 4$. The code
for a URM-program $P$ is $code(P) = [code(I_1), ..., code_d(I_d),
0]$, corresponding to a program for URCM. Clearly the programs $P$
and $code(P)$ compute the same partial function. Thus any
URM-computable function is URCM-computable. It is proved in
\cite{cut} that any recursive partial function is URM-computable,
thus is also URCM-computable. The same arguments also apply to ACM
(see \cite{boolos}).

From these results and Theorem~\ref{main-theorem} we obtain

\te{Any partial recursive function is URCM-computable and
$ACM$-computable. For any $k
> 0$, $\psi^k$ is a universal function for $k$-ary recursive
functions. }

\co{$URCM$ and  $ACM$ are universal machines (cf.
Theorem~\ref{package}). }

\section{\label{turing} Turing Machine}

The \la{Turing configuration machine (TCM)} is an arithmetic
$C$-machine defined as follows:

1. $\bb{C} = \mN \times \mN$, $\bb{S} = \bb{P} = \bb{M} = \mN$.

2. $\tau(q, u) = [q, (u)_1]$, which is primitive recursive.

3. $\bullet = \sq$.

Any number $u$ determines a two-way tape on $\mN$
\[tape(u) = (..., (u)_{0, 3}, (u)_{0, 2}, (u)_{0,1}, (u)_{0,0}, \stackrel{\triangledown}{(u)_1},
(u)_2, (u)_3, ....)\] with $\alpha(u) = (u)_0 = [(u)_{0,0},
(u)_{0,1}, ...]$ as the code for the left part (in reverse order)
and $\beta(u) = [(u)_1, (u)_2, ...]$ as the code for the right part;
$(u)_1$ is the scanned number.

A configuration $(q, u) \in \bb{C}$ consists of a state $q$ and the
code $u$ for a tape.

A situation  $[q, a] \in \bb{S}$ consists of a current state $q$ and
the scanned number $a$.

For any $(x_1, ..., x_k)$ let
\[J_k(x_1, ..., x_n) = [\underbrace{1, 1,
...,1}_{x_1 + 1}, 0, \underbrace{1, 1. ..., 1}_{x_2 + 1}, 0,
\underbrace{1, 1. ..., 1}_{x_3+1},0...]\]Each $J_k(a_1,..., a_k)$
represents the right part of the start tape when the input is $(x_1,
..., x_k)$
\[(\underbrace{\stackrel{\triangledown}{1}, 1, ...,1}_{x_1+1}, 0,
\underbrace{1, 1. ..., 1}_{x_2+1}, 0, \underbrace{1, 1. ...,
1}_{x_3+1},0...)\]

Let $\mI^k(x_1, ..., x_k) = (0, [0, J_k(x_1, ..., x_k)])$ be the
start configuration when the input is $(x_1, ..., x_n)$; here the
start state is $0$. Let $\mI = \mI^1 \cup \mI^2 \cup \mI^3 \cup
...$.

For any number $a$ let $g(a)$ be the number of appearances of $1$ in
the sequence $\{(a)_0, (a)_1, ...\}$. Let $\mO(c) = g((c)_1) +
g(\beta^2c)$; $\mO(c)$ is the number of appearance of $1$ on the
tape with the code $\beta(c)$.

One can show that $\mI^k$ and $\mO$ are primitive recursive.

To define $\circ$ we first introduce the basic acts on tapes:

Let $S_au = [(u)_0, [a, \beta^2u]]$ for any $a \ge 0$ (change the
scanned number to $a$).

Let $Ru = [[(u)_1, (u)_0], \beta^2u]$ (move the head one step to the
right)

Let $Lu = [\beta((u)_0), [(u)_{0, 0}, \beta u]]$ (move the head one
step to the left).

The acts $S_a, R, L$ have the following effects on the tape with
code $u$:
\[tape(S_au) = (..., (u)_{0, 3}, (u)_{0,2}, (u)_{0,1}, (u)_{0,0}, \stackrel{\triangledown}{a},
(u)_2, (u)_3, ....)\]
\[tape(Ru) = (..., (u)_{0,3}, (u)_{0,2}, (u)_{0,1}, (u)_{0,0}, (u)_1, \stackrel{\triangledown}{(u)_2},
(u)_3, ....)\]
\[tape(Lu) = (..., (u)_{0,3}, (u)_{0,2}, (u)_{0,1},  \stackrel{\triangledown}{(u)_{0,0}},
(u)_{1}, (u)_{2},  (u)_{3}, ....)\]

Assume $P > 0$.

An instruction $P - 1 = [q, r]$ consists of a state $q = (P - 1)_0$
and the code $r = \beta(P - 1)$ for an act on the tape. The acts $R,
L, S_a$ will be coded as $0, 1, a+2$ respectively.

Formally we define $\circ$ as follows:

If $\beta(P-1) = 0$ then $P \circ (q, u) = ((P-1)_0, Ru)$.

If $\beta(P-1) = 1$ then $P \circ (q, u) = ((P-1)_0, Lu)$.

If $\beta(P-1) = a$ for $a > 1$ then $P \circ (q, u) = ((P-1)_0,
S_{a-2}(u))$.

Then $\circ$ is primitive recursive (Definition by Cases,
cf~\cite{cut}, p.37).

Applying Theorem \ref{main-theorem} to TCM we obtain the following
theorem:

\te{Any TCM-computable function is recursive. }

A program for a Turing machine in the ordinary sense
(cf~\cite{dav2}) determines a TCM-program. It is well know that any
recursive partial function is computable by an ordinary Turing
machine (cf~\cite{boolos}~\cite{dav2}). From this and
Theorem~\ref{main-theorem} we obtain the following theorem:

\te{Any partial recursive function is TCM-computable. For any $k
> 0$, $\psi^k$is a universal function for $k$-ary recursive
functions. }

\co{$TCM$ is a universal machine (cf. Theorem~\ref{package}). }


\begin{thebibliography}{10}

\bibitem{boolos}

Boolos, G., Burgess, J., and Jeffrey, R., \emph{Computability and
Logic} (4th ed.) Cambridge UK: Cambridge University Press, (2002).

\bibitem{cut}

Cutland, N, J.,  \emph{Computability: An Introduction to Recursive
Function Theory}, Cambridge University Press, (1980).

\bibitem{dav1}

Davis, M., (ed.) \emph{The Undecidable}, Raven Press, Hewlett, NY
(1965).

\bibitem{dav2}

Davis, M.,  \emph{ Computability and Unsolvability}, McGraw-Hill
Book Company, Inc, New York (1958).

\bibitem{kleene}

Kleene, S., \emph{Introduction to Metamathematics}, North–Holland
Publishing Company, Amsterdam Netherlands, 10th impression (with
corrections of 6th reprint) (1971).

\bibitem{odi}

Odifreddi, P., \emph{Classical Recursion Theory}, North-Holland,
(1989).

\bibitem{rog}

Rogers, H. Jr., \emph{The Theory of Recursive Functions and
Effective Computability}, second edition, MIT Press (1987).

\bibitem{shoe1}

Shoenfield, J. R., \emph{Mathematical Logic}, Addison-Wesley
Publishing Co., Reading, (1967).

\bibitem{shoe2}

Shoenfield, J. R., \emph{Recursion Theory}, Berlin: Springer-Verlag,
(1993).

\bibitem{turing}

Turing, A.M., \emph{On Computable Numbers, with an Application to
the Entscheidungsproblem}, Proceedings of the London Mathematical
Society. 2 42: 230–65. (1937).

\bibitem{luo}

Luo, Z., \emph{ Clones and Genoids in Lambda Calculus and First
Order Logic}, preprint, arXiv:0712.3088v2.


\end{thebibliography}
\end{document}